\documentclass[12pt]{iopart}

\usepackage{iopams}
\usepackage{epsfig}

\newcommand{\sch}{Schr\"{o}dinger }
\newcommand{\za} {Zeldovich approximation }
\newcommand{\ada} {adhesion approximation }
\newcommand{\zb} {Zeldovich-Bernoulli }
\newcommand{\wm} {wave-mechanical }
\newcommand{\fp} {free-particle }
\newcommand{\nb} {$N$-body }
\newcommand{\n}{\noindent}

\begin{document}

\title{Wave mechanics and the \ada}
\author{C J Short and P Coles}
\address{Cripps Centre for Astronomy and Particle Theory, School of Physics and Astronomy, University of Nottingham, University Park, Nottingham, UK, NG7 2RD}
\eads{\mailto{ppxcjs@nottingham.ac.uk}, \mailto{peter.coles@nottingham.ac.uk}}

\begin{abstract} The dynamical equations describing the evolution of a
  self-gravitating fluid of cold dark matter (CDM) can be written in the form
  of a \sch equation coupled to a Poisson equation describing Newtonian
  gravity. It has recently been shown that, in the quasi-linear regime, the
  \sch equation can be reduced to the exactly solvable free-particle \sch
  equation. The \fp \sch equation forms the basis of a new approximation
  scheme -the {\it \fp approximation} - that is capable of evolving
  cosmological density perturbations into the quasi-linear regime. The \fp
  approximation is essentially an alternative to the adhesion model in which
  the artificial viscosity term in Burgers' equation is replaced by a
  non-linear term known as the {\it quantum pressure}. Simple one-dimensional
  tests of the \fp method have yielded encouraging results. In this paper we
  comprehensively test the \fp approximation in a more cosmologically relevant
  scenario by appealing to an \nb simulation. We compare our results with
  those obtained from two established methods: the linearized fluid approach and the Zeldovich approximation. We find that the \fp approximation comprehensively out-performs both of these approximation schemes in all tests carried out and thus provides another useful analytical tool for studying structure formation on cosmological scales.
\end{abstract}

\submitto{\it{J. Cosmol. Astropart. Phys.}}

\section{Introduction}

Over the last two decades a vigorous interplay between theory and
observation has led to the establishment of a standard cosmological model that
not only describes the composition and evolution of the universe as a whole,
but has also scored great successes in accounting for the large-scale
structures observed in the universe; for a recent review see, for example,
\cite{coles}. In this standard model, large-scale structure is thought to be
the result of the gravitational amplification of small density perturbations
present in the early universe; a mechanism known as gravitational
instability. The gravitational instability process is relatively easy to
understand at a qualitative level because it mostly relies only on the action
of gravity, with more complex hydrodynamical and radiative effects
playing a role only at late times when structure is already largely
developed. Moreover, when the density fluctuations are very
much smaller than the average density, the growth of fluctuations can
be handled quite accurately using first-order (linear) perturbative
techniques. On the other hand, galaxies and galaxy clusters represent density
enhancements well in excess of the mean density so we need to go beyond linear
theory if we are to study the formation of such objects in detail.

The non-linear regime of gravitational instability is analytically
intractable, except in a few cases where some special symmetry
applies. For the most part, therefore, cosmologists have been forced
to resort to numerical methods based on \nb computations in
order to study the late stages of the evolution of density
fluctuations. The use of computer simulations has revolutionized
cosmology, especially in enabling the simulation of the large-scale
structure expected to be found in galaxy redshift surveys and the
consequent testing of models using observations. In some cases,
however, such as when a very large volume and/or very high resolution
is needed, one has to make more detailed predictions than can be
obtained by using computational techniques. In such cases numerous
analytical approximations have been suggested that provide robust results in a
necessarily limited range of circumstances (e.g. see \cite{sahni, bern} and
references therein). These approximate techniques also help to promote a
genuine understanding of the principal processes involved in the formation of
cosmological structure.

Analytical attempts to understand large-scale structure formation
are commonly based on the assumption that cold dark matter (CDM) can be
treated as a self-gravitating pressureless fluid. Two important approximation
methods can be derived from the equations of motion for this fluid by applying
linear Eulerian and Lagrangian perturbation theory,
respectively. First-order Eulerian perturbation theory - the so-called {\it
  linearized fluid approach} (e.g. \cite{peeb}) - has been the backbone of
structure formation theory for many years as a result of its simplicity and
robustness at early times/on large scales. However, for Gaussian initial
conditions, it generically leads to unphysical regions where the matter
density is negative if extrapolated to late times/small scales. Although
popular, the linearized fluid approach is considerably less powerful than its
Lagrangian counterpart - the {\it \za} \cite{zel}. The Zeldovich approximation
considers (linear) perturbations in the trajectories of individual fluid
elements (particles), rather than in macroscopic fluid quantities such as the
density field. This technique is remarkably successful in describing the
pattern of structure formed from realistic (random) initial conditions
(e.g. \cite{coles2}), as long as care is taken to ensure that particle
trajectories do not intersect. Once trajectories cross (a phenomenon known as
{\it shell-crossing}), the velocity field becomes multi-valued and the density
field develops a singularity (known as a {\it caustic}). However, since the
\za is kinematical, particles do not respond to the strong gravitational
forces acting in the vicinity of the caustic. Instead they simply continue to
move along their original inertial trajectories and any structure formed is
rapidly `smeared out' in an unrealistic manner. An alternative to the \za
which avoids this problem is the adhesion model \cite{gurb}, an {\it ad hoc}
extension of the \za in which particles are assumed to `stick' to each other
when shell-crossing occurs. This sticking is achieved by including an
artificial viscosity term in the usual hydrodynamical equations of motion,
thus transforming the momentum equation into {\it Burgers'  equation}. The viscous term becomes large when particle trajectories
intersect, cancelling the component of the particle velocities perpendicular
to the caustic. This prevents particles from simply sailing through the
caustic and, consequently, the density singularities predicted by the \za are
regularized and stable structures are formed. Comparisons with \nb simulations
have shown that the adhesion model is capable of accurately reproducing the
skeleton of the `cosmic web' of filaments and voids that comprises the
morphology of the large-scale galaxy distribution \cite{kof1, wein, nuss2,
  kof2, mell}. Although the original motivation for the \ada was purely
phenomenological, it has recently been shown that Burgers' equation can be
naturally derived from the coarse-grained equations of motion, provided
certain simplifying assumptions are made \cite{buch1, buch2, dom1, dom2,
  buch3}.   

A novel alternative approach to the study of large-scale structure formation
was suggested by Widrow and Kaiser \cite{widrow}. They proposed a
wave-mechanical description of self-gravitating matter in which collisionless CDM is modelled
by a complex scalar field  whose dynamics are governed by coupled \sch and
Poisson equations (see \cite{widrow2} for a relativistic extension of the
original theory). As pointed out by Coles \cite{coles3}, the (approximately)
log-normal form of the distribution function of density perturbations can be
naturally explained within this \wm formalism. By appealing to simple
one-dimensional examples of gravitational collapse in a static universe, Coles
and Spencer \cite{coles4} showed that the \wm approach offers a competitive
alternative to the Zeldovich approximation, provided the effective Planck
constant $\nu$ in the theory is chosen carefully. They also demonstrated that,
unlike the Zeldovich approximation, the \wm approach leads to a density field
that remains non-singular when shell-crossing occurs. Building on the work of
Coles and Spencer, Short and Coles \cite{short} recently showed that, in an
expanding universe, the \sch equation can be conveniently reduced to the
\fp \sch equation in the linear and quasi-linear regimes of
gravitational instability. They advocated using the \fp \sch equation as the
basis of a new approximation scheme - the {\it \fp approximation}. The \fp
approximation is essentially an alternative to the adhesion model in which the
viscosity term is replaced by a non-linear term known as the {\it quantum
  pressure}; it is this term that prevents the formation of caustics. A convenient property of the \fp \sch equation is that, like
Burgers' equation, it possesses an analytic solution. Short and Coles \cite{short} used the \fp method to follow the gravitational evolution of a
plane-symmetric sinusoidal density fluctuation until shell-crossing. By
considering such a simple example they were able to carefully elucidate the
effect of the quantum pressure term. It was found that, if not properly
controlled, the quantum pressure term significantly impedes the growth of
density perturbations. However, in the case where the effect of this unusual
term was minimized, the \fp approximation provided an
excellent match to the \za (which is exact in one-dimension up until particle
trajectories intersect) and was capable of generating large (non-linear)
over-densities. Our objective in this paper is to provide the first
comprehensive test of this promising new method in a cosmologically relevant
scenario by appealing to a full \nb simulation. We also assess the
performance of the \fp approximation relative to the established linearized fluid approach and the Zeldovich approximation.

The layout of the paper is as follows: In section \ref{fpsect} we
present the central equations of the \fp approximation and comment on the link
between the free-particle, adhesion and Zeldovich approximations. The
essentials of the practical implementation of the \fp method are discussed in
section \ref{testfull}. In section \ref{res} we describe the results of our
comparison of the \fp approximation with an \nb simulation. We draw our
conclusions in section \ref{conc}.

\section{The \fp approximation}
\label{fpsect}

In the \fp approximation, collisionless CDM is represented by a complex scalar
field $\psi=\psi(\mathbf{x},D)$ obeying the \fp \sch equation

\begin{equation}
\label{fptdse}
\rmi\nu\frac{\partial\psi}{\partial D} = -\frac{\nu^2}{2}\nabla_{\mathbf{x}}^{2}\psi,
\end{equation}

\n where the amplitude and phase of the wavefunction are related via

\begin{equation}
\label{ampphs}
\nu\nabla_{\mathbf{x}}^{2}\left[\arg{(\psi)}\right] + \frac{1}{D}\left(|\psi|^{2}-1\right)=0,
\end{equation}

\n and the effective Planck constant $\nu$ is a free (real) parameter with
dimensions of $L^2$. For a full derivation of these equations and a discussion
of the validity of the \fp approximation, we refer the
reader to \cite{short}. The spatial coordinates $\mathbf{x}=\mathbf{x}(D)$ are
comoving coordinates and we use the linear growth factor $D=D(t)$ as a `time' variable, rather than cosmological proper time $t$. The linear growth factor is the growing mode solution of the differential equation 

\begin{equation}
\label{lg}
\ddot{D}+2H\dot{D}-4\pi G\rho_{\rm b,c}D=0,
\end{equation}

\n where a dot denotes a derivative with respect to $t$ and, at some initial
time $t_\rmi$, $D_\rmi=D(t_\rmi)=1$. The Hubble
parameter $H=H(t)$ is defined by $H\equiv \dot{a}/a$, where the scale factor
$a=a(t)$ is normalized to unity at the present epoch, and $\rho_{\rm
  b,c}=\rho_{\rm b,c}(t)$ is the density of CDM in a homogeneous
Friedmann-Robertson-Walker background cosmology. The wavefunction $\psi$ is assumed to be of the {\it Madelung form} \cite{madel}

\begin{equation}
\label{mad}
\psi=(1+\delta)^{1/2}\exp{\left(\frac{-\rmi\phi}{\nu}\right)},
\end{equation}

\n where the density contrast $\delta=\delta(\mathbf{x},D)$ is defined by
$\delta\equiv\rho/\rho_{\rm b,c}-1$ and describes fluctuations in the CDM
density field $\rho=\rho(\mathbf{x},D)$ about the homogeneous background
value $\rho_{\rm b,c}$. The gradient of the velocity potential
$\phi=\phi(\mathbf{x},D)$ gives the comoving velocity field
$\mathbf{u}=\mathbf{u}(\mathbf{x},D)$, defined by $\mathbf{u}\equiv\rmd\mathbf{x}/\rmd D$. Inserting the Madelung transformation (\ref{mad}) into (\ref{fptdse}) and (\ref{ampphs}) yields

\begin{eqnarray}
\frac{\partial\mathbf{\phi}}{\partial D}-\frac{1}{2}\left|\nabla_{\mathbf{x}}\phi\right|^2+\mathcal{P}=0,\label{fpbern}\\
\frac{\partial\delta}{\partial D}-\nabla_{\mathbf{x}}\cdot\left[(1+\delta)\nabla_{\mathbf{x}}\phi\right]=0,\label{fpcty}
\end{eqnarray}

\n and 

\begin{equation}
\label{deld}
\delta=D\nabla_{\mathbf{x}}^2\phi,
\end{equation}

\n respectively. The term $\mathcal{P}=\mathcal{P}(\mathbf{x},D)$ is known as
the quantum pressure term and is given by

\begin{equation}
\label{qp}
\mathcal{P}=\frac{\nu^{2}}{2}\frac{\nabla_{\mathbf{x}}^{2}\left[(1+\delta)^{1/2}\right]}{(1+\delta)^{1/2}}.
\end{equation}

\n The link between the \fp approximation and the more conventional Zeldovich
and adhesion approximations becomes apparent upon inspecting the Bernoulli-like
equation (\ref{fpbern}). If we were to set $\mathcal{P}\equiv 0$ then
(\ref{fpbern}) reduces to the so-called {\it Zeldovich-Bernoulli} equation
\cite{nuss} which, along with the relation (\ref{deld}), arises when the \za
is expressed in Eulerian, rather than Lagrangian, space. On the other hand,
replacing $\mathcal{P}$ by a term of the form $\mu\nabla_{\mathbf{x}}^2\phi$
leads to Burgers' equation \cite{burg} for the case of an irrotational
velocity field; this is the defining equation of the adhesion model
(e.g. \cite{gurb}). The term $\mu\nabla_{\mathbf{x}}^2\phi$ prevents
the formation of multi-stream regions and regularizes the density singularities
predicted by the Zeldovich approximation. It is controlled by the (real)
parameter $\mu$ which formally plays the role of a viscosity coefficient in
the adhesion approximation. The growth of density perturbations is suppressed
on scales $\lesssim\mu^{1/2}$ by the viscous term and thus, for large values
of $\mu$, the adhesion model leads to a much smoother distribution of matter
than observed in \nb simulations (e.g. \cite{wein}). In order to maximize the
dynamic range of the adhesion approximation, the (inviscid) limit
$\mu\rightarrow 0$ is commonly employed (e.g. \cite{gurb, kof1, kof2,
  sahni2}). In this limit, the adhesion model reduces exactly to the \za away
from regions where particle trajectories intersect. In modern approximation
methods (such as the Euler-Jeans-Newton model \cite{buch1} and the small-size
expansion \cite{dom1, dom2}) where Burgers' equation is consistently {\it
  derived} (rather than {\it assumed}), the constant viscosity coefficient
$\mu$ is replaced by a density-dependent {\it gravitational multi-stream
  coefficient} (see \cite{buch3} and references therein for a detailed
discussion).

The quantum pressure term $\mathcal{P}$ cannot be written in a form
proportional to $\nabla_{\mathbf{x}}^2\phi$ and so the \fp and adhesion
approximations are not equivalent. However, the quantum pressure term does
play a qualitatively similar role to the viscous term in the sense that it
is also a regularizing term, preventing the formation of multi-stream regions
and caustics. This was demonstrated by Coles
and Spencer \cite{coles4} who used the \sch equation to follow the evolution of a one-dimensional sinusoidal density fluctuation beyond shell-crossing. The
parameter $\nu$ controls the quantum pressure term in the same way that $\mu$
controls the viscous term and, like $\mu$, $\nu$ can also be viewed as an
approximation to a general gravitational multi-stream coefficient. Short and
Coles \cite{short} showed that, in the semi-classical limit $\nu\rightarrow
0$, the \fp approximation reduces exactly to the Zeldovich approximation prior
to shell-crossing. The quantum pressure term then only becomes important in
regions where particle trajectories intersect. However, for finite $\nu$, the
quantum pressure term has an effect before shell-crossing occurs and acts to
inhibit the gravitational collapse of density perturbations. The larger the
value of $\nu$, the greater the suppression effect; see \cite{short} for a
discussion. The \fp approximation is then no longer identical to the Zeldovich
approximation, although they may become similar at a certain distance from
collapsing regions, depending on the actual value of $\nu$. Also, once
shell-crossing occurs, the quantum pressure term has an effect outside of
multi-stream regions in the finite $\nu$ case (cf. figure \ref{fig1} of
\cite{coles4}). 

\section{Testing the \fp approximation}
\label{testfull}

In modern cosmology the process of large-scale structure formation is almost
exclusively studied using numerical \nb methods since these techniques allow
the gravitational evolution of density perturbations to be followed far into
the non-linear regime. This is something that cannot be done with any
analytical approximation. As a result, \nb simulations provide an ideal tool
for testing the applicability of analytical approximation schemes. In this
paper we test the free-particle approximation, the linearized fluid approach
and the Zeldovich approximation in Eulerian space (which we henceforth
refer to as the \zb approximation) against an \nb simulation so that we may
assess the performance of the new \fp approximation relative to existing methods. 

\subsection{The \nb simulation}

The adaptive P$^3$M code HYDRA \cite{couch} was used to perform an
\nb simulation. The simulation consisted of $N=128^3$ CDM particles contained
within a cubic box of comoving side length $L=200h^{-1}$ Mpc equipped with
periodic boundary conditions. Here $h$ is the present value of the Hubble
parameter $H_{0}$ in units of $100$ km s$^{-1}$ Mpc$^{-1}$. A spatially-flat $\Lambda$CDM
cosmological model was assumed with parameters $\Omega_{{\rm c},0}=0.24$,
$\Omega_{{\rm b},0}=0$, $\Omega_{\Lambda,0}=0.76$, $h=0.73$ where
$\Omega_{{\rm c},0}$, $\Omega_{{\rm b},0}$ and $\Omega_{\Lambda,0}$ are the
present values of the CDM, baryon and dark energy density parameters,
respectively. The simulation began at an initial scale factor $a_\rmi=0.02$ and the particles were initially positioned so as to form a random realization of a CDM density field with a power spectrum $P=P(k)$ of the form:

\begin{equation}
P = T^2 P_{\rm i},
\end{equation}

\n where $P_{\rm i}=P_{\rm i}(k)$ is the primordial power spectrum of density
fluctuations generated by inflation and $T=T(k)$ is the transfer function. We assume that the primordial power spectrum is scale-invariant:
$P_{\rm i}\propto k$, as predicted by a general class of inflationary models (see, for example, \cite{brand} for a review). The following empirical form for the CDM transfer function was adopted \cite{bard2}:

\begin{equation}
T = \frac{\ln{(1+2.34q)}}{2.34q}\left[1+3.89q+(16.1q)^2+(5.46q)^3+(6.71q)^4\right]^{-1/4},
\end{equation}

\n where $q=q(k)$ is given by $q=k\theta^{1/2}/(\Omega_{{\rm
    c},0}h^2\:\rm{Mpc}^{-1})$ and $\theta=\kappa/1.68$. Here
$\kappa=\Omega_{{\rm r},0}/\Omega_{\gamma,0}$ is the ratio of the present
relativistic particle and photon density parameters respectively; in a
universe with three relativistic neutrino flavours and photons
$\kappa=1.68$. Finally, the power spectrum was normalized so that the present
value of the rms density fluctuation in spheres of radius $8 h^{-1}$ Mpc
($\sigma_{8,0}$) was $0.74$. The Newtonian gravitational evolution of the CDM
    distribution was then followed from $a_\rmi=0.02$ to the present $a_0=1$,
    with the comoving positions of all $N$ particles stored at $20$ different
   output values of the scale factor. 

\subsection{The testing process}
\label{test}

To test the Eulerian free-particle, linearized fluid and \zb approximations
against the Lagrangian \nb simulation described above, we divide the cubic
simulation volume (of comoving side length $L=200 h^{-1}$ Mpc) into a uniform
cubic mesh with $N_{\rm g}=128^3$ grid points. There is then one particle per
grid cell (on average) and the comoving separation between neighbouring grid
points in the mesh is $\Delta=L/N_{\rm g}^{1/3}= 1.5625 h^{-1}$ Mpc, which is
roughly the size of a typical galaxy group. At each output value of the scale
factor, we use the comoving positions of the particles in the simulation to
generate the CDM density field on the cubic mesh by applying triangular-shaped
cloud interpolation (e.g. \cite{hock}). The three approximation methods we are
comparing all use the linear growth factor $D$ as a time variable, rather than
the scale factor $a$. We therefore convert the output values of the scale
factor used in the simulation to values of the linear growth factor by noting
that, in a spatially-flat $\Lambda$CDM cosmology, the solution of (\ref{lg}) is 

\begin{equation}
\label{D}
D\propto\frac{5}{6}\mathcal{B}_{\alpha}(5/6,2/3)\left(\frac{\Omega_{\rm
      c,0}}{\Omega_{\Lambda,0}a^3}\right)^{1/3}\left(1+\frac{\Omega_{\rm
      c,0}}{\Omega_{\Lambda,0} a^6}\right)^{1/2},
\end{equation} 

\n where the constant of proportionality is chosen to ensure $D_\rmi =1$,
$\mathcal{B}_\alpha$ is the incomplete beta function and $\alpha=\alpha(a)$ is defined by

\begin{equation}
\alpha=\frac{\Omega_{\Lambda,0}a^3}{\Omega_{\rm c,0}+\Omega_{\Lambda,0}a^6}.
\end{equation}

\n We now describe how the CDM density field can be calculated at any
particular value of the linear growth factor for each of the approximation
schemes we are considering.

\paragraph{The \fp approximation.}

The first step is to determine the initial velocity potential $\phi_\rmi$ on
the cubic mesh. This is done by numerically solving
$\nabla_{\mathbf{x}}^2\phi_{\rm i}=\delta_\rmi$ in Fourier space, where
$\delta_\rmi$ is the initial CDM density field of the \nb simulation. The
initial wavefunction $\psi_\rmi$ is generated on the mesh by inserting $\delta_\rmi$ and $\phi_\rmi$ into 

\begin{equation}
\label{ipsi2}
\psi_{\rm i}=(1+\delta_{\rm i})^{1/2}\exp{\left(\frac{-\rmi\phi_{\rm i}}{\nu}\right)},
\end{equation}

\n along with a suitable finite value of $\nu$. We discuss how the optimal value of
$\nu$ can be found in section \ref{nusec}. The discrete Fourier transform
$\hat{\psi}_{\rm i}=\hat{\psi}_\rmi(\mathbf{k})$ of the initial wavefunction is
then calculated and the exact solution of the \fp \sch equation (\ref{fptdse}):

\begin{equation}
\label{fpsol}
\hat{\psi}=\hat{\psi_{\rm i}}\exp{\left[\frac{-\rmi\nu (D-1)k^2}{2}\right]}
\end{equation}

\n is used to determine the Fourier transform
$\hat{\psi}=\hat{\psi}(\mathbf{k},D)$ of the wavefunction at any $D\geq 1$ of
interest. Here $\mathbf{k}$ is a comoving wavevector and $k=|\mathbf{k}|$. The
wavefunction $\psi$ in real space follows by taking the inverse discrete
Fourier transform of $\hat{\psi}$ and the CDM density field is determined
from the amplitude of the wavefunction via $\delta=|\psi|^2-1$. As we shall
see in section \ref{res}, knowledge of the velocity potential $\phi$ is also
useful. We calculate $\phi$ on the cubic mesh by using numerical Fourier techniques to solve (\ref{deld}).

\paragraph{The linearized fluid approach.}

The CDM density field $\delta$ at any $D\geq 1$ is determined from the initial
\nb density field $\delta_\rmi$ by simply applying the linear growth law $\delta=D\delta_{\rm i}$.

\paragraph{The \zb approximation.}

The initial velocity potential $\phi_{\rm i}$ is constructed on the cubic mesh as before and the \zb equation

\begin{equation}
\label{zbflat2}
\frac{\partial\phi}{\partial D}-\frac{1}{2}\left|\nabla_{\mathbf{x}}\phi\right|^2  = 0
\end{equation}

\n is integrated forwards in time from $D_{\rm i}=1$ to $D_0$ using the
the simple numerical scheme described in \cite{nuss}. At each time step in the integration we know $\phi$ and so the CDM density field $\delta$ can be numerically calculated on the mesh via (\ref{deld}).

\section{Results and discussion}
\label{res}

In order to optimize the \fp approximation, it is essential to select the
parameter $\nu$ carefully. We briefly address this issue before proceeding to
test the \fp method against the linearized fluid and \zb
approximations. First, we introduce a dimensionless parameter
$\Gamma=\nu/\Delta^2$ with $\Delta$ defined as before. The parameter $\Gamma$
emerges upon rewriting the \fp \sch equation (\ref{fptdse}) in terms of the
dimensionless comoving coordinate $\bar{\mathbf{x}}=\mathbf{x}/\Delta$. Note
that $\Gamma\propto\nu$; hereafter we will analyse the behaviour of the \fp
approximation in terms of $\Gamma$ rather than $\nu$.

\subsection{Optimizing the \fp approximation}
\label{nusec}

The numerical implementation of the \fp approximation described in section
\ref{test} requires a finite value of $\Gamma$, rather than $\Gamma\rightarrow
0$. Recall that, in this case, the quantum pressure term impedes the
gravitational collapse of density perturbations even before multi-stream
regions are formed. However, since $\mathcal{P}\propto\Gamma^2$, this
undesirable effect can be minimized by choosing the smallest possible value of
$\Gamma$. In order to find this optimal value of $\Gamma$, we first note that
the phase of the initial wavefunction (\ref{ipsi2}) is
$\arg{(\psi_\rmi)}=-\phi_\rmi/\nu\propto 1/\Gamma$. Sampling the initial phase
field at the Nyquist sampling rate requires that the change in phase between
two adjacent grid points in the cubic mesh must be less than or equal to
$\pi$ radians. As $\Gamma$ is decreased towards zero, the phase of the initial
wavefunction varies increasingly rapidly and this condition will eventually be violated. Phase-aliasing effects then cause the power spectrum of the initial
wavefunction $\psi_{\rm i}$ to become very noisy and our method breaks down. The optimal value of $\Gamma$, $\Gamma_{\rm c}$, is then defined
as the smallest value of $\Gamma$ for which there is no phase-aliasing; we
ensure there is no aliasing by using a simple numerical algorithm to
check that the phase difference between each point in the mesh and its nearest
neighbours is not greater than $\pi$ radians. The value of $\Gamma_{\rm c}$ is
mesh-dependent and, in the work presented here, $\Gamma_{\rm c}=0.12$. To
illustrate that $\Gamma=\Gamma_{\rm c}$ is indeed the optimal case, we compare
the relative sizes of the quantum pressure $\mathcal{P}$ and convective
$\mathcal{C}=-\left|\nabla_{\mathbf{x}}\phi\right|^2/2$ terms in the Bernoulli
equation (\ref{fpbern}) for several different choices of $\Gamma$. We do this
by introducing a ratio $\chi=\chi(a)$, defined by $\chi=\langle |\mathcal{P}|
\rangle/\langle |\mathcal{C}| \rangle$. For each value of $\Gamma$ we also investigate how the quantum pressure term affects the formation of large-scale structure within the \fp approximation.

\begin{figure}[htbp]
\centering
\epsfxsize=11cm
\epsfbox{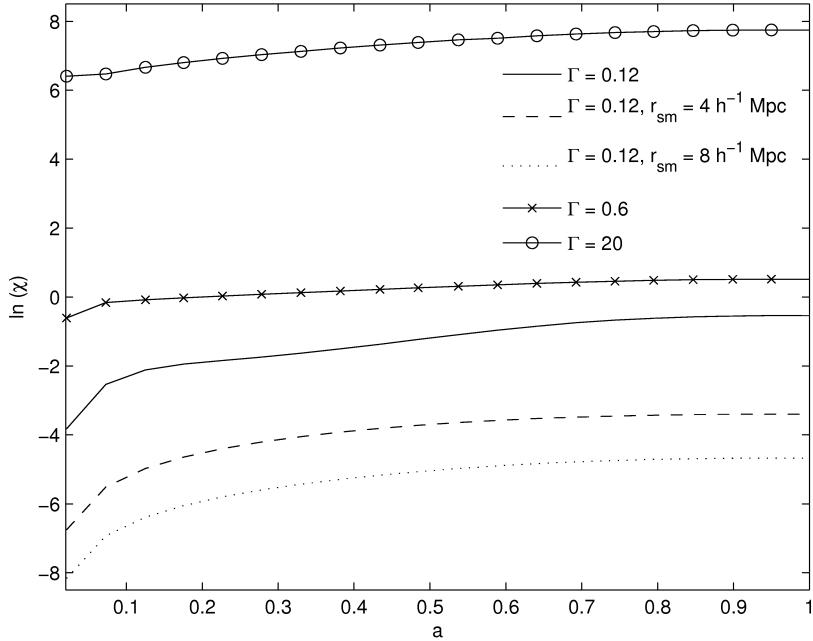}
\caption{The ratio $\chi=\langle |\mathcal{P}|\rangle /\langle
|\mathcal{C}|\rangle$ plotted as a function of the scale factor $a$ for three
different values of the parameter $\Gamma$. The dashed and dotted lines
show how $\chi$ evolves when the \fp (with $\Gamma=0.12$) density field has
been smoothed with Gaussian filters of radii $r_{\rm sm}=4h^{-1}$ Mpc and $r_{\rm sm}=8h^{-1}$ Mpc, respectively.}
\label{fig1}
\end{figure}

Figure \ref{fig1} shows how the ratio $\chi$ evolves with the scale
factor. For the moment we focus on the three solid lines, corresponding to
$\Gamma=0.12$, $\Gamma=0.6$ and $\Gamma=20$, respectively. Observe that the
ratio $\chi$ increases monotonically with the scale factor in all cases,
implying that, on average, the quantum pressure grows relative to the
convective term as time proceeds. We can also see that, at any particular
value of the scale factor, the value of $\chi$ decreases as $\Gamma$ is
decreased. Therefore, on average, the quantum pressure term becomes less
significant (relative to the convective term) as the value of $\Gamma$ is made
smaller. At each value of the scale factor, we find that the smallest value of
$\chi$ is indeed obtained for $\Gamma=\Gamma_{\rm c}=0.12$. For reasons that
will become clear in section \ref{comp}, we also examine the behaviour of
$\chi$ when the \fp density field is smoothed with a Gaussian filter prior to
calculating the quantum pressure and convective terms (note that we apply the
smoothing first to ensure numerical gradients are well behaved). The dashed
and dotted lines in figure \ref{fig1} show how $\chi$ evolves for two
smoothing scales: $r_{\rm sm}=4h^{-1}$ Mpc and $r_{\rm sm}=8h^{-1}$ Mpc,
respectively. In both cases the value of $\chi$ is very small  at all values
of the scale factor and, as the smoothing length is increased, the quantum pressure term diminishes relative to the convective  term (on average).  

\begin{figure}[htbp]
\centering
\epsfxsize=\textwidth
\epsfbox{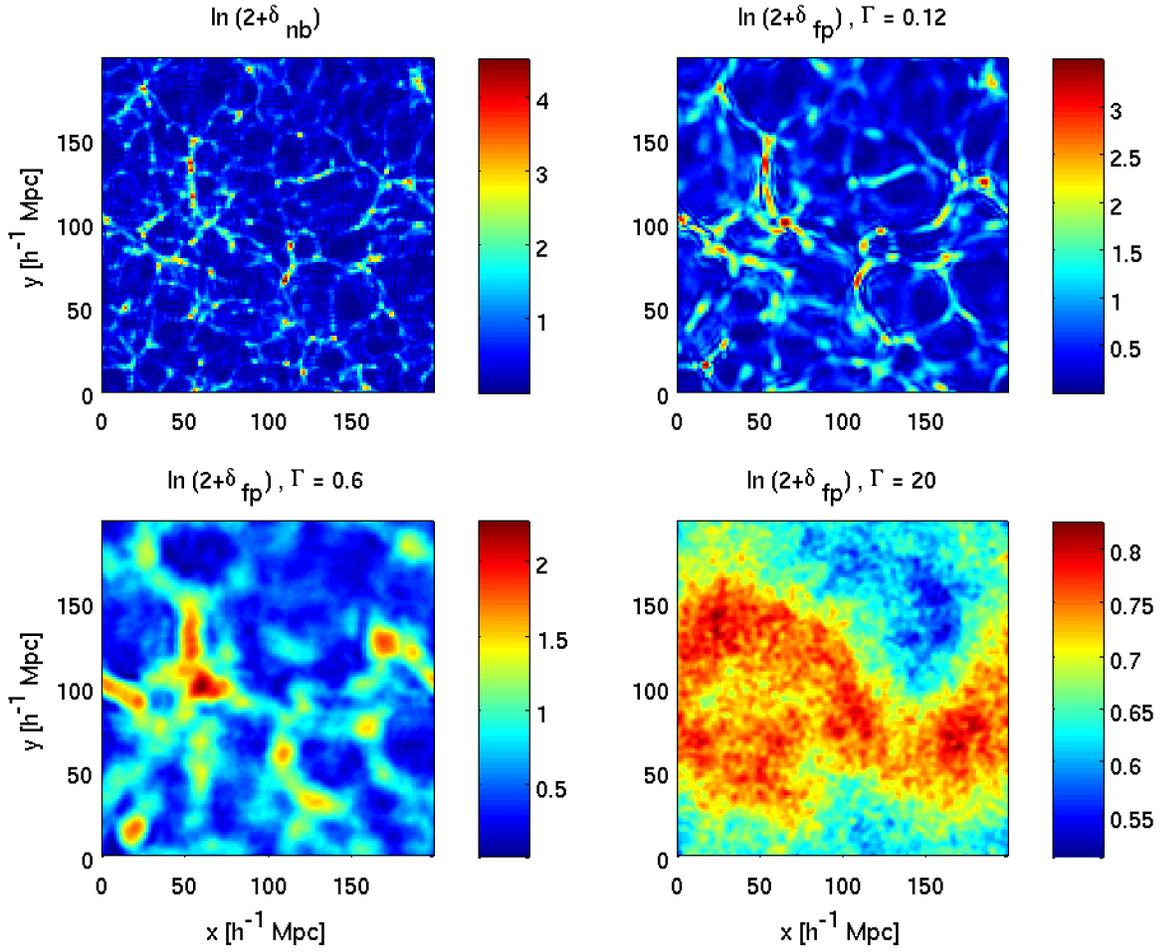}
\caption{Slices through the \nb density field $\delta_{\rm nb}$ and the \fp density fields $\delta_{\rm fp}$ for three different choices of the dimensionless parameter $\Gamma$. The value of the scale factor is $a=a_0=1$ and the slices are arbitrarily taken at a $z$-coordinate of $100h^{-1}$ Mpc.}
\label{fig2}
\end{figure}

In figure \ref{fig2} we show slices through the final (i.e. at $a=a_0=1$) \nb
and \fp density fields for the cases $\Gamma=0.12$, $\Gamma=0.6$ and
$\Gamma=20$. The global morphology of the \fp density field agrees well with
that of the \nb density field for $\Gamma=0.12$, with the peaks of the two
density fields located at the same positions. However, the over-densities
formed in the \fp approximation are not as large as in the \nb simulation and
the \fp density field is considerably smoother than the \nb density
field. This is not surprising since the \fp approximation is essentially a
quasi-linear method and thus cannot account for the full non-linear growth of
density perturbations. As the value of $\Gamma$ is increased to $0.6$ the
pattern of large-scale structure appears `smeared out' in the \fp
approximation. This is because the quantum pressure term is now larger on
average (see figure \ref{fig1}) and inhibits the collapse of density
perturbations over a greater distance; see also \cite{short}. Similar behaviour
is also seen in the \ada as the viscosity parameter $\mu$ is increased
\cite{wein}. In the case where $\Gamma=20$, the quantum pressure term
completely dominates the convective term on average (figure \ref{fig1}) and
there is no growth of density fluctuations whatsoever. Instead, the initial
density field simply oscillates rapidly in a seemingly random fashion and
remains close to homogeneous. This oscillatory behaviour was also observed in
the simple test of the \fp approximation performed by Short and Coles
\cite{short}. 

\subsection{Comparison of approximation schemes}
\label{comp}

We have established that our \fp method will be optimized for
$\Gamma=\Gamma_{\rm c}=0.12$. Accordingly, we set $\Gamma=0.12$ in the \fp
approximation for the remainder of this work. We now use the method described
in section \ref{test} to test the \fp approximation 
against an \nb simulation and compare our results with those of the linearized
fluid approach and the \zb approximation. 

\begin{figure}[htbp]
\centering
\epsfxsize=\textwidth
\epsfbox{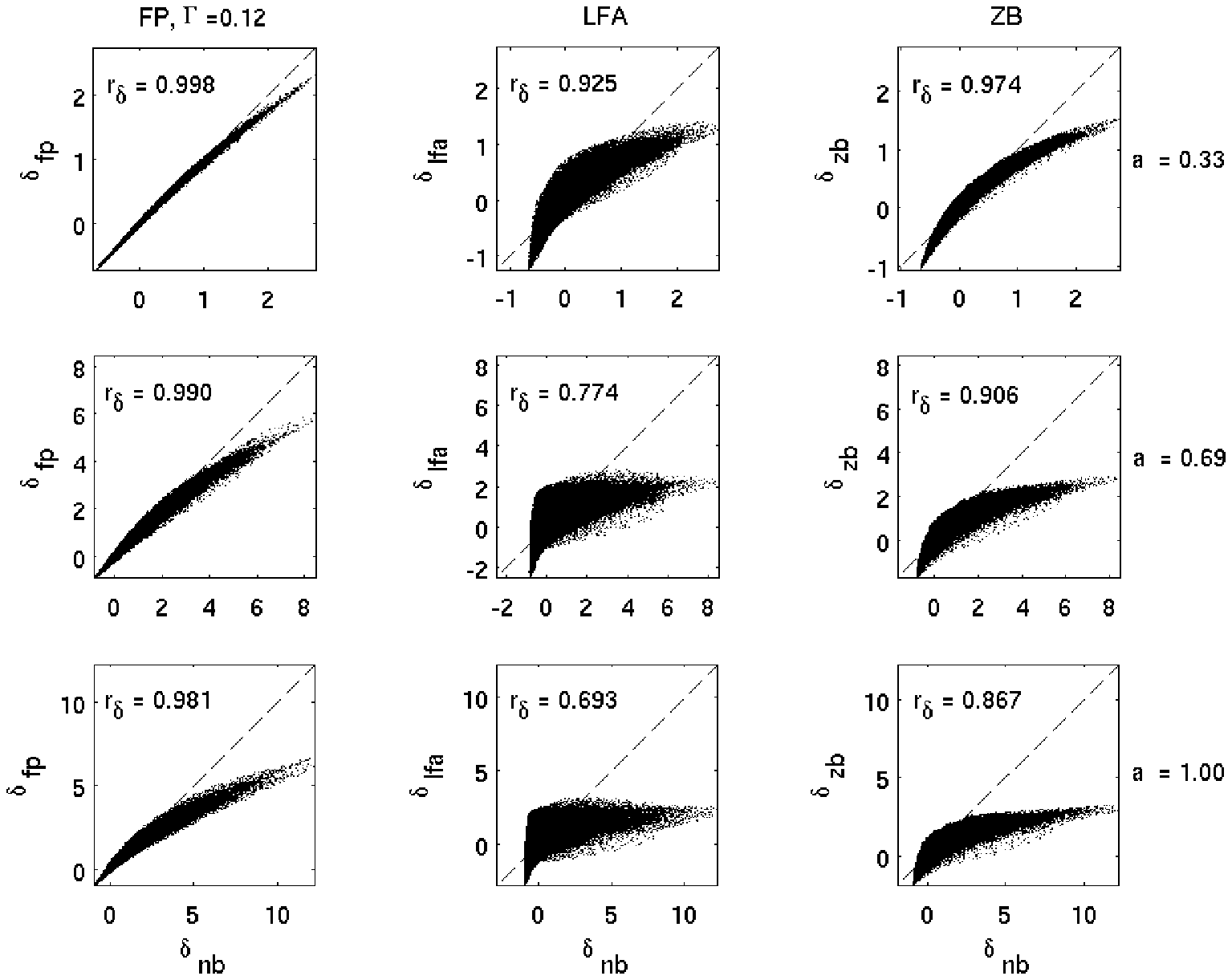}
\caption{Point-by-point comparisons of the density fields obtained from the
  three approximation schemes and the \nb density field $\delta_{\rm nb}$. The
  first, second and third columns correspond to the \fp (with
  $\Gamma=0.12$), linearized fluid and \zb approximations,
  respectively. Comparisons are shown for three different values of the scale
  factor $a$ and all of the density fields have been smoothed with a Gaussian filter of radius $r_{\rm sm}=4h^{-1}$ Mpc. The corresponding value of the correlation coefficient $r_{\delta}$ between the density fields is shown in the top-left corner of each plot.}
\label{fig3}
\end{figure}

Figure \ref{fig3} shows
point-by-point comparisons of the density fields obtained from the three
approximation schemes with the \nb density field. If the density field $\delta$ obtained from any one of our
approximation methods is identical to the \nb density field $\delta_{\rm nb}$
at all points in the cubic mesh, then there will be no scatter about the diagonal line overlaid on each plot. We quantify any scatter about the diagonal by using the correlation coefficient

\begin{equation}
\label{r}
r_{\delta} = \frac{\langle\delta_{\rm nb}\delta\rangle}{\langle\delta_{\rm nb}^2\rangle^{1/2}\langle\delta^2\rangle^{1/2}};
\end{equation}

\n a value of $r_{\delta}=1$ corresponds to the case where there is no
scatter. All of the density fields in figure \ref{fig3} have been smoothed with
a Gaussian filter of radius $r_{\rm sm}=4h^{-1}$ Mpc to smooth over highly
non-linear regions where the linear/quasi-linear approximation methods we are
testing cannot be expected to give reliable results. The main result of figure
\ref{fig3} is that the \fp approximation performs significantly better than
both the linearized fluid and \zb approximations when tested against an \nb simulation. It is clear from the point-by-point
comparisons that, at any given value of the scale factor, the \fp density
field displays the best correlation with the \nb density field. Observe that
the extrapolation of first-order Eulerian and Lagrangian perturbation theory to
late times leads to unphysical regions in the density field
where $\delta<-1$, i.e. where $\rho<0$. This problem is neatly side-stepped in
the \fp approximation since, by definition, $\delta=|\psi|^2-1$ and
$|\psi|^2\geq 0$. This explains why the \fp density field provides the best
match to the \nb density field in low density regions. It is evident from
figure \ref{fig3} that, even after applying smoothing, all of the
approximation methods systematically underestimate the density in high-density
regions. However, it is interesting to note that the \fp approximation
performs considerably better than the other approximation methods in high
density regions too.

\begin{figure}[htbp]
\centering
\epsfxsize=\textwidth
\epsfbox{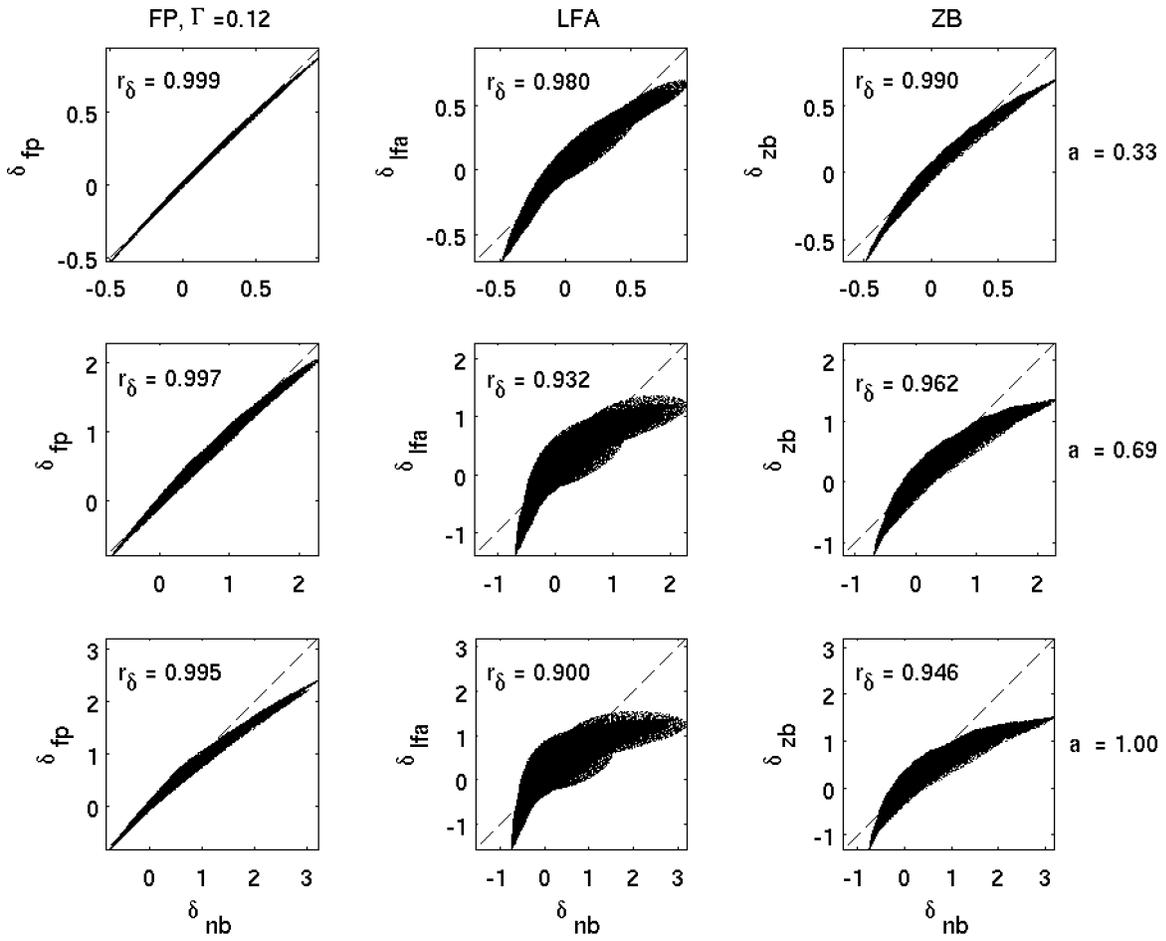}
\caption{Point-by-point comparisons of the density fields obtained from the
  three approximation schemes and the \nb density field $\delta_{\rm nb}$. The
  layout of the plots is the same as in figure \ref{fig4}; the only difference
  is that all of the density fields have been smoothed with a Gaussian filter of radius $r_{\rm sm}=8h^{-1}$ Mpc.}
\label{fig4}
\end{figure}

Figure \ref{fig4} is essentially the same as figure \ref{fig3} except that all
of the density fields have been smoothed with a Gaussian filter of radius
$r_{\rm sm}=8h^{-1}$ Mpc. By using a larger smoothing scale we restrict all of
the density fields to the quasi-linear regime $\delta\sim 1$. Consequently, the
performance of all the approximation methods is improved. However, the \fp
approximation still clearly out-performs the linearized fluid and \zb
approximations. The correlation between the smoothed \fp and \nb density
fields is very good at all values of the scale factor, with only a slight discrepancy between the two density fields in high-density
regions as $a\rightarrow 1$. Although the smoothing is now much heavier, the
linearized fluid approach and the \zb approximation still generate negative
matter densities at late times.

\begin{figure}[htbp]
\centering
\epsfxsize=11cm
\epsfbox{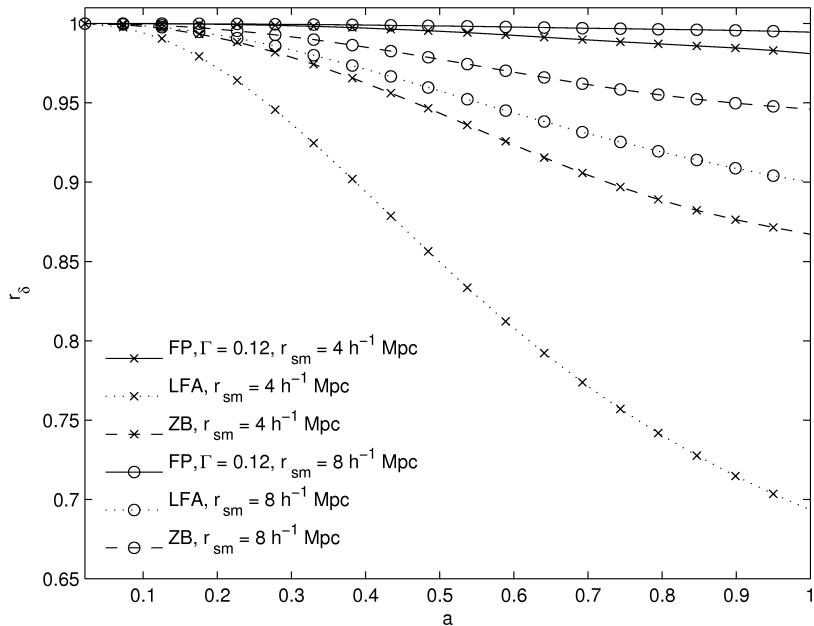}
\caption{The correlation coefficient $r_{\delta}$ for each of the
  three approximation schemes plotted as a function of the scale factor $a$. Results are shown for the two smoothing lengths $r_{\rm sm}=4h^{-1}$ Mpc and $r_{\rm sm}=8h^{-1}$ Mpc.}
\label{fig5}
\end{figure}

The correlation coefficient $r_{\delta}$ provides a simple way of
quantifying the scatter about the diagonal in the point-by-point comparisons of
figure \ref{fig3} and figure \ref{fig4}. In figure \ref{fig5} we explicitly
plot $r_{\delta}$ as a function of the scale factor for each approximation
scheme. Results are shown for the two smoothing scales: $r_{\rm sm}=4h^{-1}$
Mpc and $r_{\rm sm}=8h^{-1}$ Mpc. It is immediately clear that, for both
smoothing lengths, the \fp approximation consistently yields a higher value of
$r_{\delta}$ than the other two methods. It is interesting to note that, when
the \fp density field is smoothed with a Gaussian filter of radius $r_{\rm
  sm}=4h^{-1}$ Mpc, the value of $r_{\delta}$ obtained is always larger than
that calculated from the linearized fluid and \zb density fields smoothed on a
scale of $8h^{-1}$ Mpc. In other words, even when the linearized fluid and \zb
density fields are more heavily smoothed, the correlation with the
\nb density field is still not as good.

\begin{figure}[htbp]
\centering
\epsfxsize=11cm
\epsfbox{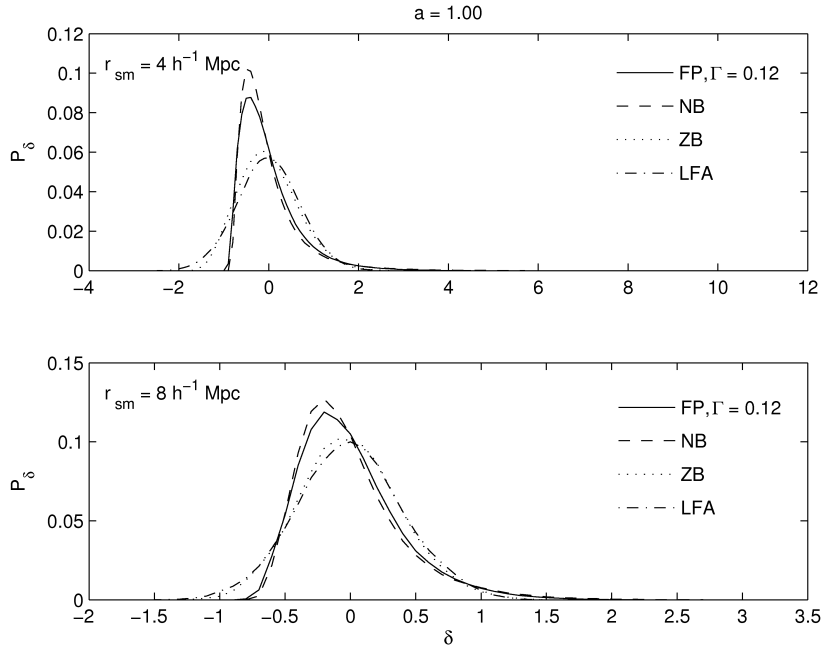}
\caption{The one-point PDFs $P_{\delta}$ of the density fields obtained from
  the \fp (with $\Gamma=0.12$), linearized fluid and \zb approximations. The
  \nb PDFs are also shown for comparative purposes. The value of the scale factor is $a=a_0=1$ and the top and bottom plots correspond to the smoothing lengths $r_{\rm sm}=4h^{-1}$ Mpc and $r_{\rm sm}=8h^{-1}$ Mpc, respectively.}
\label{fig6}
\end{figure}

A simple means of examining the statistical properties of the density
fields in the different approximation schemes is afforded by the
one-point probability distribution function (PDF)
$P_{\delta}=P_{\delta}(\delta)$ of the density fields. The PDF of the initial
\nb density field is Gaussian by construction. However, non-linear
gravitational evolution causes the \nb PDF to become non-Gaussian at late
times. The distribution becomes skewed since positive density fluctuations
(over-densities) can grow indefinitely whereas negative density
fluctuations (under-densities) cannot exceed $\delta=-1$. Figure \ref{fig6}
shows the PDFs of the various density fields at the final time $a=a_0=1$ for
the smoothing lengths $r_{\rm sm}=4h^{-1}$ Mpc and $r_{\rm sm}=8h^{-1}$
Mpc. The corresponding PDFs of the \nb density field are shown for
reference. We can see that, for both smoothing scales, the PDF of the \fp
density field provides the best match to the \nb PDF for all values of
$\delta$. The fact that the \fp approximation guarantees $\delta\geq 1$ causes
the \fp PDF to be skewed in the same way as the \nb PDF. It is clear from both
plots in figure \ref{fig6} that the PDF of the linearized fluid density field
is Gaussian. This must be the case for Gaussian initial conditions since the
density contrast at a given point simply grows proportional to the linear
growth factor to first-order in Eulerian perturbation theory. The PDFs of the
\zb density field also seem to be close to Gaussian. Consequently, the
linearized fluid and \zb approximations both assign a non-zero probability to the existence of regions with $\delta <-1$ at late times.

\begin{figure}[htbp]
\centering
\epsfxsize=11cm
\epsfbox{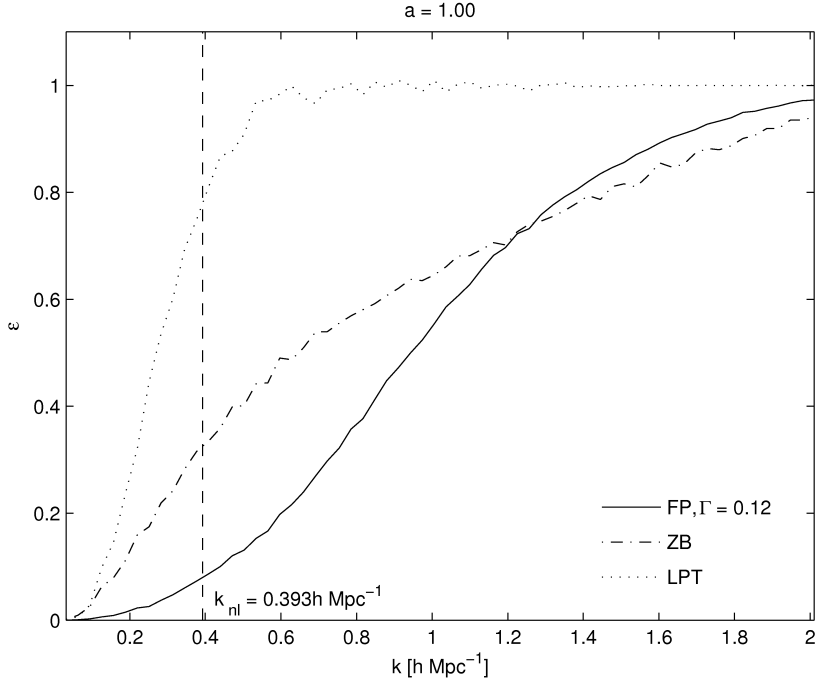}
\caption{The statistic $\epsilon$ plotted as a function of the comoving
  wavenumber $k$ for each of the three approximation schemes. The value of the
  scale factor is $a=a_0=1$. The dashed vertical line corresponds to the
  wavenumber $k_{\rm nl}=0.3927h$ Mpc$^{-1}$ which indicates the boundary between Fourier modes in the linear and non-linear regimes.}
\label{fig7}
\end{figure}

In our discussion so far we have only compared the performance of the
free-particle, linearized fluid and \zb approximations in real space. However, it is informative to examine the behaviour of the different approximation
schemes in Fourier space as well. For each method we investigate how the
Fourier components $\hat{\delta}=\hat{\delta}(\mathbf{k},a)$ of the
appropriate density field evolve relative to the corresponding Fourier
components $\hat{\delta}_{\rm nb}=\hat{\delta}_{\rm nb}(\mathbf{k},a)$ of the
\nb density field. To do this we use the statistic $\epsilon=\epsilon(k,a)$, defined by

\begin{equation}
\label{epsi}
\epsilon  = \frac{\sum |\hat{\delta}-\hat{\delta}_{\rm nb}|^2}{\sum (|\hat{\delta}|^2+|\hat{\delta}_{\rm nb}|^2)},
\end{equation}

\n where, for each (comoving) wavenumber $k$, the summations are over all
wavevectors $\mathbf{k}$ with a magnitude in the interval
$(k-2\pi/L,k]$. The quantity $\epsilon$ provides a measure of the
differences between the amplitudes and phases of the Fourier components
$\hat{\delta}$ and $\hat{\delta}_{\rm nb}$. If the amplitudes and phases of the
Fourier components of the two fields are identical then $\epsilon=0$. On the
contrary, if the phases of the two fields are completely uncorrelated then
$\epsilon=1$ on average. A convenient property of the statistic $\epsilon$ is
that it is independent of any smoothing of the density fields and so we are
free to use the un-smoothed density fields to calculate
$\epsilon$. Figure \ref{fig7} shows $\epsilon$ as a function of the wavenumber
$k$ at the final time $a=a_0=1$ for the different approximation schemes we
are considering. The vertical dashed line shown in the plot denotes the
wavenumber $k_{\rm nl}=0.3927h$ Mpc$^{-1}$ which corresponds to a wavelength of
$\lambda_{k_{\rm nl}}=16h^{-1}$ Mpc. From observations, the present rms
density fluctuation in spheres of diameter $\lambda_{k_{\rm nl}}$ is of
order unity and so the scale $k_{\rm nl}$ can be taken as an indication of the
boundary between modes in the linear and non-linear regimes of gravitational
evolution. Fourier modes with a wavenumber $k<k_{\rm nl}$ are still in the
linear regime and evolve independently of each other. The phase information of
the initial density field is preserved in this case. On the other hand, modes
with a wavenumber $k\geq k_{\rm nl}$ experience phase shifts due to coupling
between different modes in the non-linear regime (e.g. \cite{ryden,
  chiang}). It is apparent from figure \ref{fig7} that, for all of the
approximation schemes, the value of $\epsilon$ becomes larger as $k$ is
increased. This is not surprising since the various methods we are comparing
cannot accurately follow the non-linear evolution of small-scale modes. The
main point of figure \ref{fig7} is that, for wavenumbers in the range
$0<k\lesssim 3k_{\rm nl}$, the value of $\epsilon$ is consistently smaller in
the \fp approximation than in the linearized fluid and \zb approximations. In
particular, the \fp approximation performs much better than the other two
methods for $0<k<k_{\rm nl}$. Notice that the \zb approximation gives a
slightly lower value of $\epsilon$ than the \fp approximation for $k\gtrsim
3k_{\rm nl}$. However, the wavenumber $k=3k_{\rm nl}$ corresponds to a
wavelength $\lambda_{3k_{\rm nl}}=5.3333h^{-1}$ Mpc and we find that the rms
density fluctuation in spheres of diameter $\lambda_{3k_{\rm nl}}$ is
approximately $2$ for the final \nb density field. Thus we cannot
realistically expect the \fp and \zb approximations to yield reliable results on scales much smaller than this anyway.

\section{Conclusion}
\label{conc}

In this paper we have investigated a new approach to the study of cosmological
structure formation known as the \fp approximation. The \fp
approximation is similar in spirit to the \ada in the sense that it is an
extension of the \za which includes a regularizing term - the quantum pressure
term - to prevent the formation of density singularities when shell-crossing
occurs. The quantum pressure is controlled by a free parameter $\nu$ (or
$\Gamma$ in the notation of section \ref{res}) which is akin to the viscosity
coefficient $\mu$ in the adhesion model. In the semi-classical limit
$\nu\rightarrow 0$ the quantum pressure only has an effect in multi-stream
regions. However, in this work we have employed a numerical implementation of
the \fp approximation that uses a finite value of $\nu$ instead. In this case
the quantum pressure term acts to suppress the collapse of density
perturbations away from regions where particle trajectories cross; the degree
of suppression increases with $\nu$.

The \fp method has been rigorously tested by appealing to a full cosmological
\nb simulation. We have shown that the performance of the \fp approximation,
relative to the \nb simulation, is optimized when $\nu$ is set to the smallest
possible value $\nu_{\rm c}$ supported by the discrete mesh we use. In this
case the effect of the quantum pressure is minimal. For larger values of
$\nu>\nu_{\rm c}$ the quantum pressure term becomes more significant relative
to the convective term in (\ref{fpbern}) and the details of the large-scale
structure distribution appear `washed out' in the \fp approximation. In
particular, for $\nu\gg\nu_{\rm c}$, the quantum pressure term is dominant and
initial density fluctuations undergo oscillation rather than growth. The \fp
approximation (with $\nu=\nu_{\rm c}$) consistently yields better results than
the established linearized fluid and \zb approximations in all of the tests
discussed in section \ref{res}. At all times the density field obtained from
the \fp method is considerably better correlated with the \nb density field
than the density fields of the other two approximation schemes. This is
particularly true in low density regions which is a consequence of the fact
that the \fp density field is guaranteed to be non-negative. This feature of
the \fp approximation also explains why the one-point PDF of the \fp density
field provides the closest match to the \nb PDF of all the approximation
methods. We have also seen that the \fp approximation out-performs the
linearized fluid approach and the \zb approximation in Fourier space. From a
practical point of view, another advantage of the \fp approximation is that it
is very quick to implement. This is because the \fp \sch equation has an exact
solution, allowing us to directly calculate the density and velocity fields at
any given time without having to numerically integrate equations of motion. 

To conclude, the \fp approximation provides a quick and effective way of
evolving cosmological density perturbations into the quasi-linear regime in any
CDM-dominated cosmology. We have compared the \fp method with two
traditional approaches and the results have proved favourable. In light of
these comments, we believe that the \fp approximation provides another useful
addition to the repertoire of analytical techniques available for the study of
large-scale structure formation.

\ack

C J Short would like to thank PPARC for the award of a studentship that made
this work possible. Also, we are grateful to F R Pearce for many useful
comments regarding numerical issues and to the anonymous referee for helpful suggestions.

\end{document}